\documentclass[aps,prb,twocolumn,superscriptaddress,raggedbottom,showpacs,floatfix]{revtex4-1}
\usepackage{amsmath,amsfonts,amssymb,amsthm}
\usepackage{bm}
\usepackage{color}
\usepackage{graphicx,latexsym}
\usepackage{epstopdf}

\begin{document}
\renewcommand{\vec}{\mathbf}
\renewcommand{\Re}{\mathop{\mathrm{Re}}\nolimits}
\renewcommand{\Im}{\mathop{\mathrm{Im}}\nolimits}

%\title{ Topological Kondo insulator beyond the mean-field approximation: a strongly-interacting Dirac liquid on the surface}
\title{A Strongly-Interacting Dirac Liquid on the Surface of a Topological Kondo Insulator}
\author{Dmitry K. Efimkin}
\affiliation{Joint Quantum Institute and Condensed Matter Theory Center, Department of Physics, University of Maryland, College Park, Maryland 20742-4111, USA}
\author{Victor Galitski}
\affiliation{Joint Quantum Institute and Condensed Matter Theory Center, Department of Physics, University of Maryland, College Park, Maryland 20742-4111, USA}

\begin{abstract}
A topological Kondo insulator (TKI) is a strongly-correlated material, where hybridization between the conduction electrons and localized $f$-electrons gives rise to a crossover from a metallic behavior at high temperatures to a topologically non-trivial insulating state at low temperatures. The existing description of the TKIs is based on a  slave-boson mean-field theory, which neglects  dynamic fluctuation phenomena. Here, we go beyond the mean-field theory and investigate the role of Kondo fluctuations on the topological surface states. We derive an effective theory of the Dirac surface states coupled to fluctuations and show that the latter mediate strong repulsive interactions between surface excitations. We show that these effects renormalize the plasmon spectrum on the surface. % This fluctuation-induced ``straightening'' of the plasmon dispersion is strongly temperature dependent, which provides an experimental tool to distinguish the non-trivial fluctuation effects from more conventional plasmon dynamics.
 We also argue that Kondo-mediated interactions may drive a magnetic instability of the surface spectrum.  
\end{abstract}
\maketitle

{\em Introduction --} The last decade has witnessed a flurry of interest and tremendous advance in the studies of topological states of matter. By now, a new class of topological insulating state has been theoretically predicted~\cite{KaneMele1,KaneMele2,MooreBalents,FuKaneMele,BernevigHughesZhang,FuKane} and experimentally discovered in both two- and three-dimensional material systems (see Refs.~[\onlinecite{HasanKane, QiZhang,KonigBuhmannMolenkampHughes}] and references therein). Moreover a complete topological classification of non-interacting band structures has been put together~\cite{SchnyderRyuFurusakiLudwig,Kitaev}. Despite this impressive progress, there still remain a number of open fundamental questions, and understanding the interplay between interactions and topological states is arguably the chief challenge  among them on the theoretical side~\cite{LuVishwanath,WangPotterSenthil}.

While some of these fundamental questions are academic at this stage, there is also an experimental motivation to look closer at the effect of interactions on the topological surface states, which comes from the recent discovery of a topological Kondo insulator in Samarium hexaboride~\cite{Paglione1,Exp1,Exp2,Exp3,Exp4,Exp5,Exp6,Exp7,Exp8,Exp9,Exp10,Exp11}. Indeed, these heavy-fermion topological insulators represent the first known class of material where the topological behavior arises not from the non-interacting band structure, but due to strong interactions and correlations~\cite{TKIGalitskiPRL,TKIGalitskiPRB}. Furthermore, a recent experiment by the Paglione group~\cite{Paglione2} on $\mathrm{SmB_6}$ suggests that interactions in this TKI are not a minor complication, but may be of the utmost importance, because they appear to drive the topological surface states into a ferromagnetic phase. The nature of the latter is not  understood as of now and this calls for a critical revision of the underlying theoretical framework.

The theoretical description of a Kondo insulator is usually based on the Anderson model [see, Eq.~(\ref{AH}) below], which describes hybridization between the  conduction $d$-electrons and localized, strongly-interacting $f$-electrons. To solve the model,  the slave boson technique \cite{Coleman1} is usually employed, which allows to conveniently circumvent complications associated with strong (formally, infinite) repulsion between the $f$-electrons by ``splitting'' the physical $f$-electron into a product of a fermion and a slave boson, supplemented with a constraint to remove the double occupancy. A mean-field analysis (reiterated in more detail below) involves condensing the boson field and neglecting all of its dynamics. This effectively leads to a non-interacting model of an insulator, where the gap is proportional to the condensate  and hybridization parameter, and makes it amenable to a topological analysis. Since the electron states being hybridized have the opposite parities, the resulting Kondo insulator is topological~\cite{TKIGalitskiPRL,TKIGalitskiPRB} and contains its hallmark feature - the Dirac surface states.

There is a good evidence that the mean-field theory is reasonable, and actually works at least at the qualitative level.  In particular, the existing data on $\mathrm{SmB_6}$, including the observation of surface states there \cite{Exp1,Exp2,Exp5}, appear to be in a good agreement with it. However, there are also reasons, both experimental and theoretical, to study effects beyond mean field and this is the focus of our work. 

There has been a large amount of prior work on the effect of Kondo fluctuations in heavy-fermion materials both metallic~\cite{Coleman2,MillisLee, Millis, MillisLavagnaLee,AuerbachLevin,ReadNewns1,ReadNewns2,Read} and insulating~\cite{Karbowski}. The theory is well-controlled in the large-$N_\mathrm{f}$ limit (where $N_\mathrm{f}$ is the degeneracy of the $f$ level) and the mean-field approximation becomes exact for $N_\mathrm{f} \to \infty$. In realistic systems (with $N_\mathrm{f}=2$ or $N_\mathrm{f}=4$), the consensus appears to be that fluctuations  play an important role in heavy-fermion metals, but are essentially irrelevant in (conventional) Kondo insulators. The case of a topological Kondo insulator considered here is of a mixed character, because the ``matter fields,'' which the Kondo fluctuations are coupled to, are gapped in the bulk, but gapless on the surface. 

To succinctly summarize the results of our theory detailed below, we find that Kondo fluctuations  manifest themselves as strong short-range repulsive temperature-dependent interactions between surface Dirac states. Therefore, a proper description of the TKI's surface state is a {\em strongly-interacting Dirac liquid.} The interaction can drive the magnetic instability on the surface, which is accompanied by opening of an insulating gap in the Dirac spectrum. In the absence of a phase transition, these interactions distort the spectrum of surface collective modes.

{\em Mean-field slave-boson theory of a TKI --}
Our starting point is the following Anderson lattice Hamiltonian   
\begin{equation}
\begin{split}
\label{AH}
\hat{H}=\sum_{ij \sigma} \epsilon^{\mathrm{d}}_{ij} \hat{d}_{{i} \sigma}^\dagger \hat{d}_{j\sigma} +\sum_{ij \alpha} \epsilon^{\mathrm{f0}}_{ij}\hat{f}_{{i}\alpha}^\dagger \hat{f}_{{j}\alpha} + \\
+ \sum_{ij\sigma\alpha}( V^{\sigma \alpha}_{ij} \hat{d}_{i\sigma}^\dagger \hat{f}_{j\alpha} + \mathrm{h.c.}) + \frac{U}{2}\sum_{i\alpha} \hat{f}_{i\alpha}^\dagger \hat{f}_{i\alpha} \hat{f}_{i\bar{\alpha}}^\dagger \hat{f}_{i\bar{\alpha}},
\end{split}
\end{equation}
where the first two terms describe hopping of  $d$- and $f$-electrons, with $\epsilon^{\mathrm{d}}_\vec{k}$ and $\epsilon^{\mathrm{f}0}_\vec{k}$ being their bare dispersion laws as a function of the lattice momentum, ${\bf k}$. As our goal is to extract new physics associated with the  fluctuation effects in TKIs, we would like to separate them from non-universal phenomena associated with  a particular compound. Hence, we will focus from now on  a generic topological Kondo insulator with the simplest band structure and assume that the relevant $f$-states form a Kramers doublet (labeled by the index $\alpha=\uparrow, \downarrow$ and $\bar{\alpha}=\downarrow,\uparrow$),  $\epsilon^{\mathrm{d}}_\vec{k}=-E_\mathrm{d}+k^2/2m_\mathrm{d}$ and $\epsilon^{\mathrm{f}0}_\vec{k}=-E_\mathrm{f}^0-k^2/2m_\mathrm{f}$. The model requires an ultraviolet cutoff, $E_\mathrm{d}^\mathrm{m}$, which corresponds to the width of the conduction band.  All energies are measured from the position of the unoccupied $f$-levels.  As shown below, the resulting theory is described by $4\times 4$ matrices in the bulk and one Dirac cone on the surface (as opposed to a $8\times 8$ bulk Hamiltonian and  three surface Dirac cones for $\mathrm{SmB_6}$~\cite{AlexandrovDzeroColeman,LuZhaoWengFangDai,LegneRueggSigrist}). The third term in Eq.~(\ref{AH}) describes  hybridization between the $f$- and $d$-states. Since they have different parities, which dictates $\hat{V}_{-\vec{k}}=-\hat{V}_\vec{k}$ for their Fourier transforms, and the time reversal symmetry is respected, the hybridization matrix element at small momenta can be approximated as $\check{V}_\vec{k}=V \vec{k} \cdot \check{\bm{\sigma}}$. This particular form of the hybridization is the source of non-trivial topology of the emergent bands.  The last term in Eq.~(\ref{AH}) is the Hubbard repulsion between two $f$ electrons on a level.  The Hubbard energy scale, $U$, much exceeds all other energy scales in the problem and hence the last term effectively enforces a  no-double-occupancy constraint on each site. This constraint is ultimately a source of fluctuation effects we are studying.

To exclude the double occupation of the $f$-states we use the slave-boson approach, with spinon $\hat{f}_i$ and holon $\hat{b}_i$ operators introduced in the following way
\begin{equation}
\hat{f}_{i\alpha}\rightarrow \hat{f}_{i\alpha} \hat{b}_i^\dagger, \quad \hat{f}_{i\alpha}^\dagger\rightarrow \hat{f}_{i\alpha}^\dagger \hat{b}_i. 
\end{equation}
Singly-occupied $f$-states correspond to the occupation numbers  $|1_{\uparrow\mathrm{s}},0_{\downarrow\mathrm{s}},0_\mathrm{h}\rangle$ ($|0_{\uparrow\mathrm{s}},1_{\downarrow\mathrm{s}},0_\mathrm{h}\rangle$) in terms of the spinons and holons, and an empty $f$-level corresponds to a state occupied by a holon $|0_{\uparrow\mathrm{s}},0_{\downarrow\mathrm{s}},1_\mathrm{h}\rangle$. The doubly-occupied $f$-levels  are excluded by the constraint on each site $\hat{Q}_i=0$ (operator identity), where $\hat{Q}_i$ is given by 
\begin{equation}
\hat{Q}_i=\sum_{\alpha}\hat{f}_{i\alpha}^\dagger \hat{f}_{i\alpha} + \hat{b}_i^\dagger \hat{b}_i-1.  
\end{equation}
The reason behind this construction is that an equality is straightforward to enforce in the Lagrangian formalism by introducing a functional delta-function (while the original inequality in terms 
of the physical $f$-electrons is not as easy to implement). The Lagrangian for our system takes the form:
\begin{equation}
\begin{split}
\label{L}
L=\sum_{i}\bar{b}_i(\tau)\partial_\mathrm{\tau}b_i(\tau) 
+ \sum_i \bm{i} \lambda_i(\tau) Q_i(\tau) + \\ +
\sum_{ij}\bar{\Psi}_{i} (\tau) \left(\begin{array}{cc} \partial_\tau+\epsilon^\mathrm{d}_{ij} & \check{V}_{ij} \bar{b}_j(\tau) \\ b_i (\tau) \check{V}_{ij}^\dagger &\partial_\tau +b_i (\tau)\epsilon^\mathrm{f0}_{ij}\bar{b}_j (\tau)  \end{array}\right)\Psi_{j} (\tau).
\end{split}
\end{equation}
 Where $\Psi_i=(d_{i\uparrow},d_{i\downarrow},f_{i\uparrow},f_{i\downarrow})^{\rm T}$ is a column of Grassmann fields combining the $d$- and $f$-electrons, and $\lambda_i(\tau)$ is a time-dependent auxiliary real field, which enforces the no-double-occupancy constraint. $\check{V}$ is a $2\times 2$ hybridization matrix in the spin/pseudo-spin space and the diagonal components of the matrix in (\ref{L}) are proportional to a $2\times 2$ identity matrix (not shown). % Also here it needs to be born in mind that $\bar{f}_i b_i \bar{b}_i f_i\equiv \bar{f}_i f_i$ due to the constrain.

The phase degree of freedom of slave particles is a redundant variable and can be integrated out after a gauge transformation~\cite{ReadNewns1,ReadNewns2,Read}, which separates the phase, $\theta_i$, and amplitude of holon field $b_i^\mathrm{a}$  
\begin{equation}
b_i=b_i^\mathrm{a} e^{\bm{i} \theta_i}, \quad f_i \rightarrow f_i e^{\bm{i} \theta_i}, \quad \lambda_i\rightarrow \lambda_i - \frac{d \theta_i}{d \tau}. 
\end{equation}
In mean field theory, the holon field condenses and it, together with the  auxiliary field, acquires site- and time-independent values $b_i^{\mathrm{a}}\rightarrow b_0$ and $\quad \bm{i}\lambda_i\rightarrow \delta E_\mathrm{f}$. The resulting mean-field Hamiltonian, $\hat{H}_{\mathrm{MF}}$, reads
\begin{equation}
\label{MFHam}
\hat{H}_{\mathrm{MF}}=\sum_\vec{k} \hat{\Psi}_\vec{k}^\dagger\left(\begin{array}{cc} \epsilon^\mathrm{d}_\vec{k} & b_0 \check{V}_\vec{k} \\ b_0 \check{V}_\vec{k}^\dagger &\epsilon^\mathrm{f}_\vec{k}  \end{array} \right) \hat{\Psi}_\vec{k},
\end{equation}
where $\epsilon^\mathrm{f}_\vec{k}=-E_\mathrm{f}^0+\delta E_\mathrm{f} - b_0^2 k^2/2 m_\mathrm{F}$ is a renormalized dispersion law of the $f$-states. The spectrum consists of two bands $\epsilon^{\pm}_\mathrm{\vec{k}}$, separated by direct gap $\Delta_\vec{k}=b_0(\mathrm{Tr}[\check{V}_\vec{k}^\dagger \check{V}_\vec{k}]/2)^{1/2}$ and given by   
$
\epsilon^\pm_\vec{k}=\left(\epsilon_\vec{k}^\mathrm{d}+\epsilon_\vec{k}^\mathrm{f}\right)/2 \pm \sqrt{\left(\epsilon_\vec{k}^\mathrm{d}-\epsilon_\vec{k}^\mathrm{f}\right)^2/4+\Delta_\vec{k}^2}
$.
There is an indirect gap $\Delta_{\mathrm{b}}\sim \Delta_\vec{k}^2/E_\mathrm{d}^\mathrm{m}$ and it is assumed that the chemical potential of TKI bulk lies within it. The set of self-consistent equations for $b_0$ and $\delta E_\mathrm{f}$ can be derived from the mean-field action and are given by
\begin{equation}
b_0^2+\sum_{\alpha}\langle \hat{f}_{i\alpha}^\dagger \hat{f}_{i\alpha} \rangle =1, \quad b_0 \delta E_\mathrm{F}=-\sum_{i\sigma\alpha}V_{ij}^{\sigma\alpha}\langle \hat{f}_{j\alpha}^\dagger \hat{d}_{i\sigma} \rangle.  
\end{equation}
These equations should be suplemented by additional one for the number of particles, $n_\mathrm{f}+n_\mathrm{d}=N_\mathrm{f}$, which imposes full occupation of the band $\epsilon_{\vec{k}}^{-}$. This equation fixes the position of the chemical potential for renormalized energy bands. The position of the chemical potential, and Fermi momentum are changing within gap opening, that can be expected from Luttinger's theorem arguments \cite{MillisBook,Nozieres1,Nozieres2,Otsuki,Burdin}	. The phase diagram of the model, depending on the energy of localized states, degeneracy and the lattice symmetry, has been extensively discussed recently~\cite{TKIGalitskiPRB,TranTakimotoKim,AlexandrovDzeroColeman,LegneRueggSigrist,MDzero,AlexandrovColeman}. Here we are interested in studying Kondo fluctuations relative to the mean-field saddle point. So, we treat the mean-field values, $b_0$ and $\delta E_\mathrm{f}$, as parameters of the model and assume that the system is within TKI phase. Note the following hierarchy of energy scales: $\Delta_\mathrm{b}\ll \Delta_\vec{k} \ll E^\mathrm{m}_\mathrm{d}$. The shift of the $f$-levels is of order $\delta E_\mathrm{f}\sim E_\mathrm{d}^\mathrm{m}$ and the Kondo temperature where the formation of an insulating band structure commences can be estimated as $T_\mathrm{K}\sim \Delta_\mathrm{b}$. The holon condensate $b_0$ appears at temperatures of order $\Delta_\vec{k}$, and smoothly increases with decreasing of temperature. %and $b_0\ll 1$. 

{\em Kondo fluctuations in the TKI's bulk --}
Here we restore the dynamics to the fluctuations of holon condensate $\delta b^\mathrm{a}_i(\tau) = b_i^\mathrm{a} (\tau) - b_0$ and the constraint field, $\delta \lambda_i(\tau) = \lambda_i(\tau) + {\bf i} \delta E_\mathrm{f}$. A Lagrangian, describing Gaussian Kondo fluctuations  and their interactions with electrons, is given by 
\begin{equation}
\label{Lfl1}
\begin{split}
L_{\mathrm{fl}}^\mathrm{b}=\sum_{ij}\bar{\Psi}_{i}(\tau) \left(\begin{array}{cc} 0 & \check{V}_{ij} \delta b_j^\mathrm{a}(\tau) \\ \delta b_i^\mathrm{a}(\tau) \check{V}_{ij}^\dagger & \bm{i} \delta \lambda_{j}(\tau)  \end{array}\right)\Psi_{j}(\tau) +  \\ + \sum_i\left\{ \delta E_\mathrm{f} \left[ \delta b_i^\mathrm{a}(\tau)\right]^2 + 2 {\bf i} b_0 \delta b_i^\mathrm{a}(\tau) \delta \lambda_i(\tau)  \right\}.
\end{split}
\end{equation}
The ``valence fluctuations'' of the auxiliary field, $\lambda_i$ (describing local fluctuations of $f$-states' occupation) do not have a bare mass or bare field-theory all together, and therefore its bare fluctuations are gapless. Note that an effective field theory for $\delta \lambda_i$ appears upon integrating out the matter fields - the $f$-electrons. On the other hand, the fluctuations of the holon amplitude are massive and can be integrated out. Introducing Fourier components for fermionic and bosonic fields, we get a continuous low-energy part of the fluctuation Lagrangian 
\begin{equation}
\label{Lfl2}
L_{\mathrm{fl}}^\mathrm{b}=\sum_{\vec{k}\vec{q}} {\bf i} \lambda_{\vec{q}}(\tau) \bar{f}_{\vec{k}+\vec{q}\alpha}(\tau) f_{\vec{k}\alpha}(\tau)+\sum_{\vec{q}} \frac{|\delta \lambda_{\vec{q}}(\tau)|^2}{2 U_\mathrm{b}},
\end{equation}
Here $U_\mathrm{b}=\delta E_\mathrm{f}/2b_0^2 n_\mathrm{b}$ corresponds to short-range repulsive interactions mediated by Kondo-fluctuations and $n_\mathrm{b}$ is the three-dimensional concentration of $f$-levels. It is very strong since energy of interaction between two electrons on a cite is of order $\delta E_\mathrm{f}/b_0^2$ and considerably exceeds the gap, $\Delta_\mathrm{b}$. The value of the interaction decreases with the formation of the holon condensate. It can be shown that the interaction $U_\mathrm{b}$ is inverse-proportional to the degeneracy $N_\mathrm{f}$ of the $f$-states. So in the limit, $N_\mathrm{f}\rightarrow \infty$, the slave-boson  mean-field theory is exact. For a finite $N_\mathrm{f}$, the strong short-range interaction is important in both thermodynamics and transport if bulk is metallic as was extensively discussed \cite{Coleman2,MillisLee, Millis, MillisLavagnaLee,AuerbachLevin}. But if the bulk is insulating, electronic excitations at low temperatures $T\ll \Delta_\mathrm{b}$ form a dilute gas, for which short-range interaction is unimportant. A TKI has an insulating bulk but topologically protected surface states. These Dirac surface states inherit interactions mediated by the Kondo fluctuations in the bulk. 

{\em The Dirac surface states --} Now we consider a TKI with open boundary conditions (requiring that the electron's wave-function vanish on the surface), and find the topological surface states from the mean-field Hamiltonian (\ref{MFHam}). If the surface is perpendicular to the $z$-axis and the TKI occupies the $z<0$ half-space, the contribution of surface states to the field operator $\hat{\Psi}_\mathrm{s}(\vec{r},z)$ is given by    
\begin{equation}
\hat{\Psi}_\mathrm{s}=\sum_{\vec{p}} \psi(z) e^{i \vec{p} \vec{r}} \left[\hat{\chi}_{\vec{p}}^\uparrow\left(\begin{array}{c}  u\\0\\iw\\0\end{array} \right)  + \hat{\chi}_{\vec{p}}^{\downarrow} \left(\begin{array}{c} 0\\u\\0\\-iw\end{array} \right) \right].
\end{equation}
Here $\hat{\chi}_{\vec{p}}^{\uparrow, \downarrow}$ are a pair of annihilation operators for surface states with the in-plane momentum, $\vec{p}$, and up-/down-spin. Projection of the bulk mean-field Hamiltonian onto these states yields the Dirac Hamiltonian $\check{\cal H}_\mathrm{s}=v_\mathrm{s} [\vec{p}\times \bm{\check\sigma}]_z$ with the velocity $v_\mathrm{s}=2uw b_0 V\approx 2 b_0^2 V \sqrt{m_\mathrm{d}/m_\mathrm{f}}$, which is temperature dependent.  The coherence factors $u^2$ and $w^2$ define contributions to the surface states from the conduction and $f$-states and are given by $u^2=b_0^2 m_\mathrm{d}/( m_\mathrm{f} + b_0^2 m_\mathrm{d})$ and $w^2=m_\mathrm{f}/(m_\mathrm{f} + b_0^2 m_\mathrm{d})$. Since $m_\mathrm{d}\ll m_\mathrm{f}$, then $u^2\ll w^2$ and Dirac surface states mostly consist of the $f$-electron states. We find the wave function of the surface states exponentially decaying in TI's bulk and vanishing on its surface as follows $\psi(z)=N \left[ e^{\eta_- z}-e^{\eta_+ z} \right]$. Here $N$ is a normalization factor, and $\eta_\pm$ are given by

\begin{equation}
\eta_\pm=\sqrt{\kappa_1-\kappa_2 \pm \sqrt{(\kappa_1-\kappa_2)^2-\kappa_2^2}},
\end{equation}
with $\kappa_1=2 m_\mathrm{d}m_\mathrm{f} V^2/\hbar^2$ and  $\kappa_2=2 m_\mathrm{d}m_\mathrm{f} (E_\mathrm{d}-E_\mathrm{f})/\hbar (m_\mathrm{f}+b_0^2 m_\mathrm{d})$.

The full field operator $\hat{\Psi}=\hat{\Psi}_\mathrm{s}+\hat{\Psi}_\mathrm{b}$ has contributions from both the surface, $\hat{\Psi}_\mathrm{s}$, and the interior, $\hat{\Psi}_\mathrm{b}$. To project the bulk model onto the surface we insert $\hat{\Psi}$ into the action  (\ref{Lfl1}) and neglect terms that contain the bulk states, $\hat{\Psi}_\mathrm{b}$. This approximation is justified, if the energy of surface states is deep inside the gap of the bulk spectrum. Neglecting variation of the fluctuation field in the transverse direction (on the length-scale of order penetration length of the surface states), integrating the action over $z$, and integrating out massive fluctuations of the holon field's amplitude, $\delta b_i^\mathrm{a}$, we obtain the continuous low-energy model for surface Dirac states 

\begin{equation}
\label{FlSurfaceAction}
\begin{split}
L_\mathrm{s}(\tau)=\sum_{\vec{p}} \bar{\chi}_{\vec{p}}(\tau) (\partial_\tau + v_\mathrm{s} [\vec{p}\times \check{\bm \sigma}]_z-\mu)\chi_\vec{p}(\tau) +  \\
+ \sum_{\vec{p},\vec{q}} i w^2 \lambda_{\vec{q}}(\tau) \bar{\chi}_{\vec{p}+\vec{q}}(\tau)\chi_\vec{p}(\tau) + \frac{w^4}{2 U_\mathrm{s}}\sum_{\vec{q}} |\lambda_\vec{q}(\tau)|^2.  
\end{split}
\end{equation}
Here $\chi_\vec{p}=(\chi_{\vec{p}}^\uparrow,\chi_{\vec{p}}^\downarrow)^{\rm T}$ and $\mu$ is the chemical potential of surface states; $U_\mathrm{s}=\delta E_\mathrm{f} w^4/2 b_0^2 n_\mathrm{s}$ is the Fourier transform of an effective short-range interaction mediated by the Kondo fluctuations; $n_\mathrm{s}$ is the surface concentration of $f$-sites. As seen from Eq.~(\ref{FlSurfaceAction}), the valence fluctuations mediate repulsive interactions between the emergent Dirac excitations in the density-density channel, whose strength depends on the properties of the underlying mean-field state. With the formation of the holon condensate, Kondo fluctuations and interections mediateed by them are decreasing.

Now we discuss possible manifestations of these fluctuations/interactions in the observables. Since our theory is generic we can not provide universal predictions, but we can discuss
two distinct scenarios: either interactions lead to a phase transition or not.

{\em Modification of the plasmon spectrum --}
\begin{figure}
\label{Fig1}
\begin{center}
\includegraphics[width=8.5 cm]{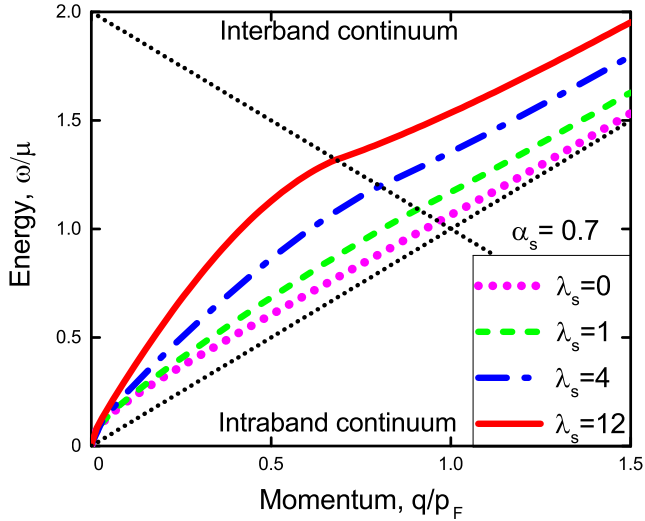}
\caption{(Color online)  Dispersion law of plasmons on the surface of a TKI,  calculated from (\ref{RPA}), for dimensionless Coulomb coupling constant $\alpha_\mathrm{s}=e^2/\hbar v_\mathrm{s} \epsilon =0.7$ and for different values of dimensionless short-range  coupling constant $\lambda_\mathrm{s}=\nu_\mathrm{F}U_\mathrm{s}=0, 1, 4, 12$. Black dotted lines are borders of continuums  $\omega<v_\mathrm{s}q$ and $\omega>2 \mu-v_\mathrm{s} q$, corresponding to intraband and interband single-particle excitations.}
\end{center}
\end{figure}
If the effective repulsion is sufficiently weak and no phase transition takes place, Kondo fluctuations do not induce a reconstruction of the surface spectrum, but have only qualitative consequences (much like in the theory of Landau Fermi-liquid). 
One such effect is a correction to the spectrum of collective modes on the surface. Since the two-dimensional Dirac electrons are charged, there is a long-range Coulomb interaction, which gives rise to collective charge oscillations -- plasmons. This Coulomb interaction between Dirac surface states can be introduced by inserting an additional term in the action (\ref{FlSurfaceAction}) 
\begin{equation}
L_{\phi}=\sum_{\vec{p},\vec{q}} i \phi_{\vec{q}}(\tau) \bar{\chi}_{\vec{p}+\vec{q}} (\tau) \chi_\vec{p}(\tau) + \sum_{\vec{q}} \frac{|\phi_\vec{q}(\tau)|^2}{2 U_\vec{q}}.  
\end{equation}
Here $\phi_\vec{q}(\tau)$ is a bosonic field, which originates from the Hubbard-Stratonovich decoupling of the Coulomb interaction and plays the role of a scalar electrical potential; $U_\vec{q}=2\pi e^2/\epsilon q$ is the Fourier transform of the  Coulomb potential in two dimensions and $\epsilon$ is an effective dielectric permittivity of TKI surface. %Both Coulomb interaction and Kondo-fluctuation-mediated interactions are in charge-charge channel. 
Fields $\phi_\vec{q}$  and $\lambda_\vec{q}$ can be incorporated in a single field $\Phi_\vec{q}=\phi_\vec{q}+w^2 \lambda_\vec{q}$ and one of them is redundant and can be integrated out. After the integration the action for $\Phi_\vec{q}$ is given by    
\begin{equation}
L_{\Phi}=\sum_{\vec{p},\vec{q}} i \Phi_{\vec{q}}(\tau) \bar{\chi}_{\vec{p}+\vec{q}} (\tau) \chi_\vec{p} (\tau)  + \sum_{\vec{q}} \frac{|\Phi_{\vec{q}}(\tau)|^2}{2(U_\vec{q}+U_\mathrm{s})}.
\end{equation}
In the random-phase approximation (RPA), the dispersion law of collective excitations satisfies the equation
\begin{equation}\label{RPA}
1-(U_\vec{q}+U_\mathrm{s})\Pi(\omega,\vec{q})=0,
\end{equation}
where $\Pi(\omega,\vec{q})$ is the polarization operator of a Dirac electron gas, which has been calculated in Refs.~[\onlinecite{WunchStauberSolsGuinea}] and [\onlinecite{HwangDasSarma}] (See also Ref.~[\onlinecite{PlasmonReview}] and references therein). The dispersion law of collective excitations depends on two dimensionless coupling constants  $\alpha_\mathrm{s}=e^2/\hbar v_\mathrm{s} \epsilon$ and $\lambda_\mathrm{s}= \nu_\mathrm{F} U_\mathrm{s}$, which describe the strength of the interactions. Here $\nu_\mathrm{F}=\mu/2\pi (\hbar v_\mathrm{s})^2$ is the density of Dirac states at the Fermi level. For $v_\mathrm{s}q\ll\omega\ll \mu$ the polarization operator can be approximated as $\Pi(\omega,\vec{q})=\nu_\mathrm{F}(v_\mathrm{s}q)^2/2\omega^2$.  In the long-wave-length limit, the long-range Coulomb interaction dominates, but at $q\gtrsim q_0=p_\mathrm{F}\alpha/\lambda$, where $p_\mathrm{F}$ is the Fermi momentum, the short-range interaction takes over. For $q\lesssim q_0$, the plasmon dispersion is the standard in two dimensions square-root law, $\omega=v_\mathrm{s}\sqrt{\alpha_\mathrm{s} p_\mathrm{F}q/2}$, while for $q_0\lesssim q\ll p_\mathrm{F}$ it is a linear zero-sound-like mode, $\omega=v_\mathrm{s}\sqrt{\lambda_\mathrm{s}/2} \cdot q$.  For estimates, we have used the following set of parameters, relevent to $\hbox{SmB}_6$ \cite{AlexandrovDzeroColeman,Exp2}, $v_\mathrm{s}\approx0.5\times 10^5\; \hbox{m}/\hbox{s}$, $\delta E_\mathrm{f}\approx 0.15\;\hbox{eV}$, $b_0\approx0.05$, $\epsilon\approx 50$, $\Delta_\mathrm{b}\approx1 \;\hbox{meV}$, $n_\mathrm{s}\approx4\times 10^{14}\;\hbox{cm}^{-2}$, which corresponds to the coupling constants $\alpha_\mathrm{s}\approx 0.7$ and $\lambda_\mathrm{s}\approx 1.3$. The dispersion law of surface plasmons for $\alpha_\mathrm{s}=0.7$ and different values of parameter $\lambda_\mathrm{s}=0, 1, 4, 12$ is presented in Fig.~1. If $\omega<v_\mathrm{s}q$ or $\omega>2 \mu-v_\mathrm{s} q$  the plasmons are strongly damped due to intraband (Landau damping) or interband transitions (the corresponding borders are shown by  dotted lines in the figure). Since plasmons are undamped only for $q<p_\mathrm{F}$, the fluctuation-mediated interaction is important in that case only if $q_0\ll p_\mathrm{F}$, which is satisfied if $\alpha_\mathrm{s}\ll\lambda_\mathrm{s}$. 

{\em Magnetic phase transition in the Dirac liquid --} While the gapless Dirac spectrum on a TKI's surface is protected by the time reversal symmetry, this symmetry can be spontaneously broken due to sufficiently strong interactions (with the appearance of an out-of-plane spin polarization~\cite{BaumStern1} or spin-density wave~\cite{BaumStern2} order parameter). An out-of-plane spin polarization opens a gap in the Dirac spectrum, which can be energetically favorable since it reduces both the kinetic energy and exchange interaction energy of Dirac electrons. This state can also be interpreted as an excitonic insulator of electrons and holes from the conduction and valence bands of the surface Dirac spectrum, the possibility of which has been widely discussed in the context of  graphene~\cite{Exciton1,Exciton2,Exciton3}.  The critical value of short-range repulsion is~\cite{BaumStern1} $U_\mathrm{s}^*=4\pi v_\mathrm{F}\hbar/k_{\Lambda}$, where $k_{\Lambda}\approx\Delta_\mathrm{b}/\hbar v_\mathrm{F}$ is an ultraviolet cutoff of the linear spectrum and $\Delta_\mathrm{b}$ is the  indirect gap in TKI's bulk. Using the same set of parameters as above, we get $U_\mathrm{s}/U_\mathrm{s}^*\approx 0.6$ and conclude that the short-range repulsion mediated by Kondo fluctuations is in the neighborhood of the critical value in realistic materials and may potentially drive the system into a magnetic phase. Furthermore, since the interaction strength is decreasing with lowering  temperature, a highly unusual scenario of a double magnetic transition (where a ferromagnetic state would appear as an intermediate phase in a finite temperature window) is conceivable. Apart from this speculative scenario, another possibility is that the Dirac liquid on the TKI's surface remains magnetically ordered in the ground state, which may be consistent with the experiment of the Paglione group~\cite{Paglione2} in Samarium hexaboride (where, however, the situation is complicated by the presence of three Dirac cones, whose Dirac points are offset relative to one another \cite{AlexandrovDzeroColeman,LuZhaoWengFangDai,RoyGalitski}).

%To conclude, we have shown that Kondo fluctuations in af TKI mediate strong repulsive temperature dependent interaction between the Dirac states on its surface. So proper description of surface states is strongly interacting Dirac Ferni liquid. The interaction can can drive the magnetic instability of the surface spectrum which is accompained by the gap opening in the surface spectrum. 

This research was supported by DOE-BES DESC0001911 (VG), DARPA (DE), and Simons Foundation. The authors are grateful to Maxim Dzero, Johnpierre Paglione, and Bitan Roy for useful discussions.

\bibliographystyle{apsrev}
\bibliography{TKIBibliography}

\end{document}